\documentclass{sf2a-conf2018}
\usepackage{graphicx}
\usepackage{hyperref}
\usepackage[]{natbib}  
\usepackage{epstopdf}

\def\BibTeX{{\rm B\kern-.05em{\sc i\kern-.025em b}\kern-.08em
    T\kern-.1667em\lower.7ex\hbox{E}\kern-.125emX}}
\bibpunct{(}{)}{;}{a}{}{,}  

\begin{document}

\TitreGlobal{SF2A 2018}

\title{Waves in the radiative zones of rotating, magnetized stars}
\author{A. Valade}\address{ENS de Lyon, 46 all\'ee d'Italie, 69007 Lyon, France}\secondaddress{AIM, CEA, CNRS, Universit\'e Paris-Saclay, Universit\'e Paris Diderot, Sorbonne Paris Cit\'e, F-91191 Gif-sur-Yvette, France}
\author{V. Prat$^2$}
\author{S. Mathis$^2$}
\author{K. Augustson$^2$}

\setcounter{page}{237}

\maketitle

\begin{abstract}
    Asteroseismology has reached a level of accuracy that may allow us to detect the effect of a deep magnetic field on oscillation modes.
    We thus aim to develop an asymptotic theory for short-wavelength waves in radiative zones of rotating stars with a general magnetic field topology (toroidal and poloidal).
    A dispersion relation is derived for a uniformly rotating star with a low-amplitude, axisymmetric magnetic field.
    The parameter space of this model is explored with a Hamiltonian ray-tracing method.
    The features of the gravito-inertial waves modified by the magnetic field are studied with different magnetic topologies.
\end{abstract}


\section{Introduction}

As of today, no direct observational technique allows us to directly probe the interior of a star.
However, asteroseismology provides us with constraints on the internal structure and rotation of stars.
Similarly, in this work we aim to characterise magneto-gravito-inertial waves (whose restoring forces are the Lorentz, buoyancy, and Coriolis forces) in order to be able to assess the magnitude and morphology of the internal magnetic field of stars that show gravity-like oscillations.
We focus here on rapidly rotating stars, which mostly concerns intermediate-mass and massive pulsators such as $\gamma$ Doradus, $\delta$ Scuti, $\beta$ Cephei, SPB, or Be stars, but the obtained results can also be applied to sub- and red giant stars \citep[e.g.][]{Fuller, LoiPapaloizou}.

\section{Derivation of the dispersion relation}

We consider a compressible, non-dissipative fluid inside a uniformly rotating star deformed by the centrifugal acceleration.
After linearising the governing magneto-hydrodynamical equations, we make the Cowling approximation, and we also neglect the derivatives of background quantities to simplify the calculations.
This leads to a linear differential system for velocity fluctuations.
Eventually, we make the WBKJ approximation, which allows us to transform the system into a matrix equation.
The dominant terms of the determinant of this matrix yield the dispersion relation for magneto-gravito-inertial waves:
\begin{equation}
  \label{eq:magi_disp}
  (\omega^2 - \omega_{\rm A}^2)^2 - \left(N_0^2 \frac{ k_\perp^2}{k^2}  +
  f^2 \frac{k_z^2}{k^2} \right) (\omega^2 - \omega_{\rm A}^2) - f^2 \frac{k_z^2}{k^2} \omega_{\rm A}^2 = 0,
\end{equation}
where $\omega$ is the angular frequency of the wave, $\omega_{\rm A}=\vec k\cdot\vec B_0/\sqrt{\rho_0\mu_0}$ is the Alfv\'en frequency, $\vec B_0$ is the background magnetic field, $\mu_0$ is the vacuum permeability, $f = 2\Omega$, $\Omega$ is the rotation rate, $N_0$ is the Brunt-V\"ais\"al\"a frequency defined by $N_0^2 = (\vec\nabla\log\rho_0 - \frac{1}{\Gamma_1} \vec\nabla\log P_0) \cdot \vec g_0$, $P_0$, $\rho_0$ and $\vec g_0$ are respectively the background pressure, density and gravity vector, $\Gamma_1$ is the first adiabatic exponent, and $k$, $k_\perp$ and $k_z$ are respectively the norm, the meridional component orthogonal to gravity and the component along the rotation axis of the wavevector $\vec k$.
A similar relation has previously been found by \cite{mathis2011} for a toroidal magnetic field. 

Two different families of waves emerge from this relationship, gravito-inertial waves modified by
the presence of a magnetic field, and Alfv\'en waves modified by the stratification and the rotation
of the star. 

\section{Ray-tracing}

To explore the properties of these waves we have used a ray-tracing method.
This approach considers waves as Hamiltonian perturbations that propagate following the group velocity at constant frequency.

Derivatives of background quantities are needed in the dispersion relation to correctly describe the behaviour of waves near the surface.
Since they are missing in Eq.~(\ref{eq:magi_disp}), we modified it by adding terms responsible for the back-refraction of gravito-inertial waves in \cite{prat2016}.
Furthermore, we only worked with magnetic fields confined inside the star to prevent waves from escaping the star.

\begin{figure}[]
    \begin{center}
        \hfill
        \includegraphics[height=.2\textwidth]{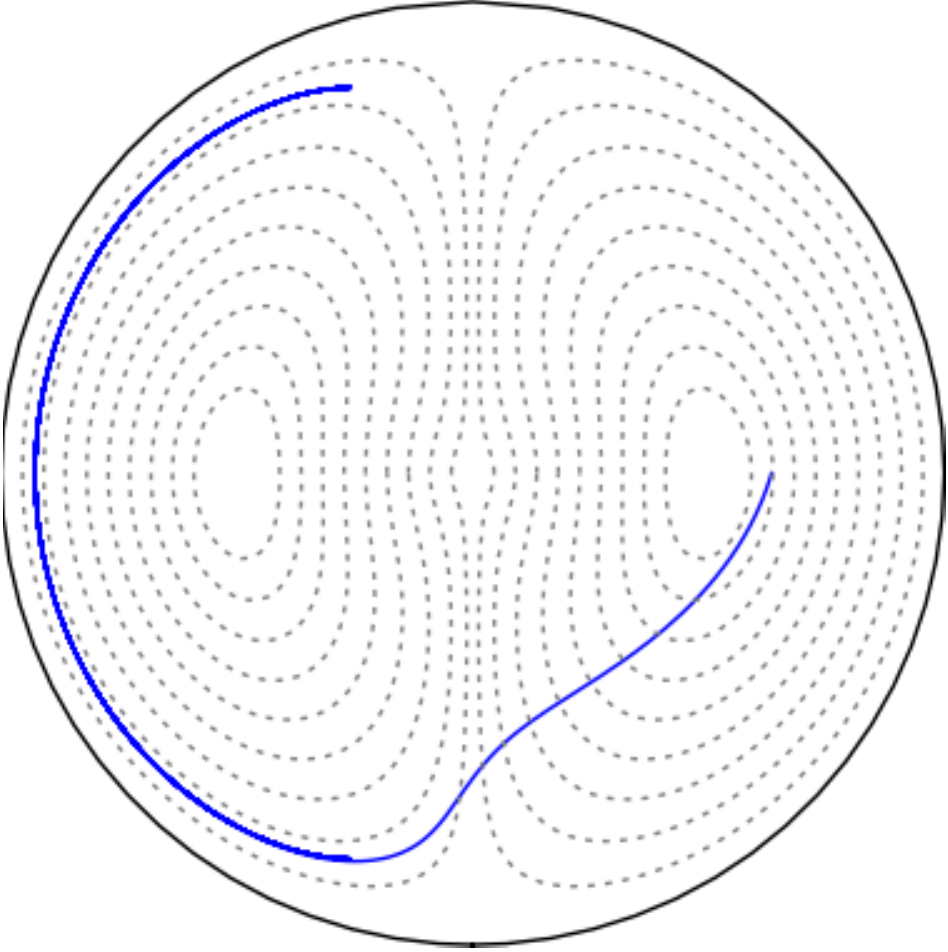}
        \includegraphics[height=.2\textwidth]{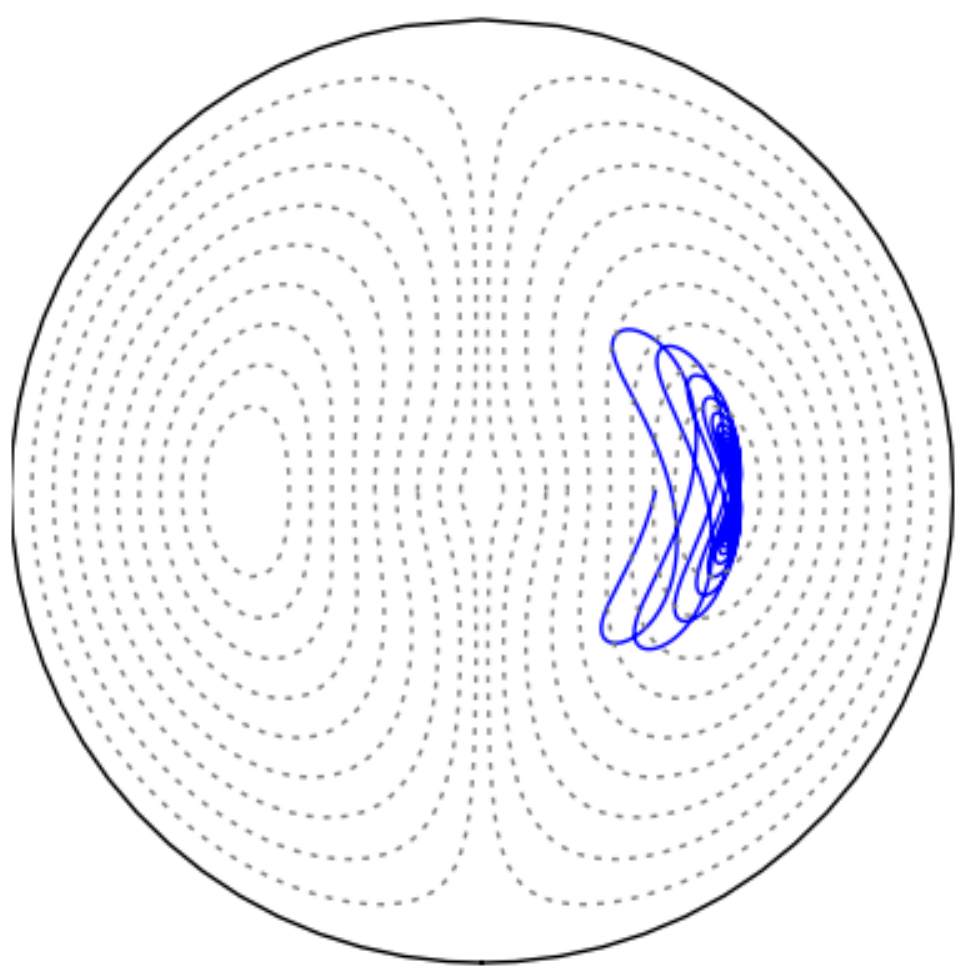}
        \hfill
        \includegraphics[height=.2\textwidth]{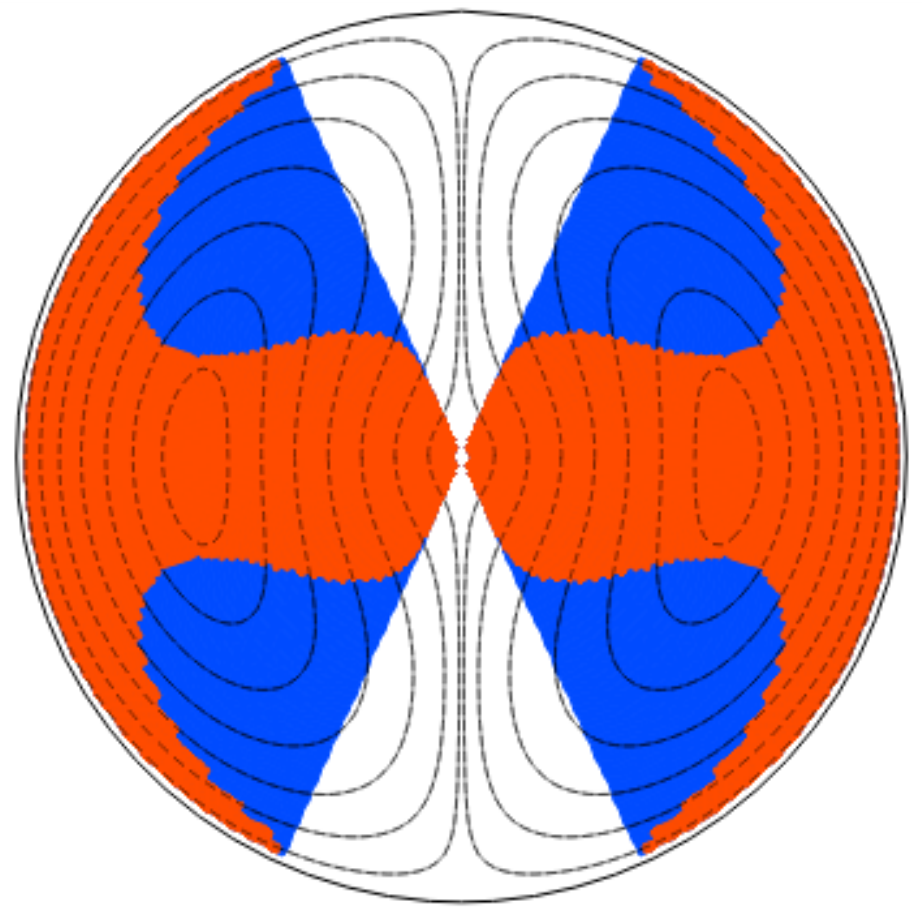}
        \includegraphics[height=.2\textwidth]{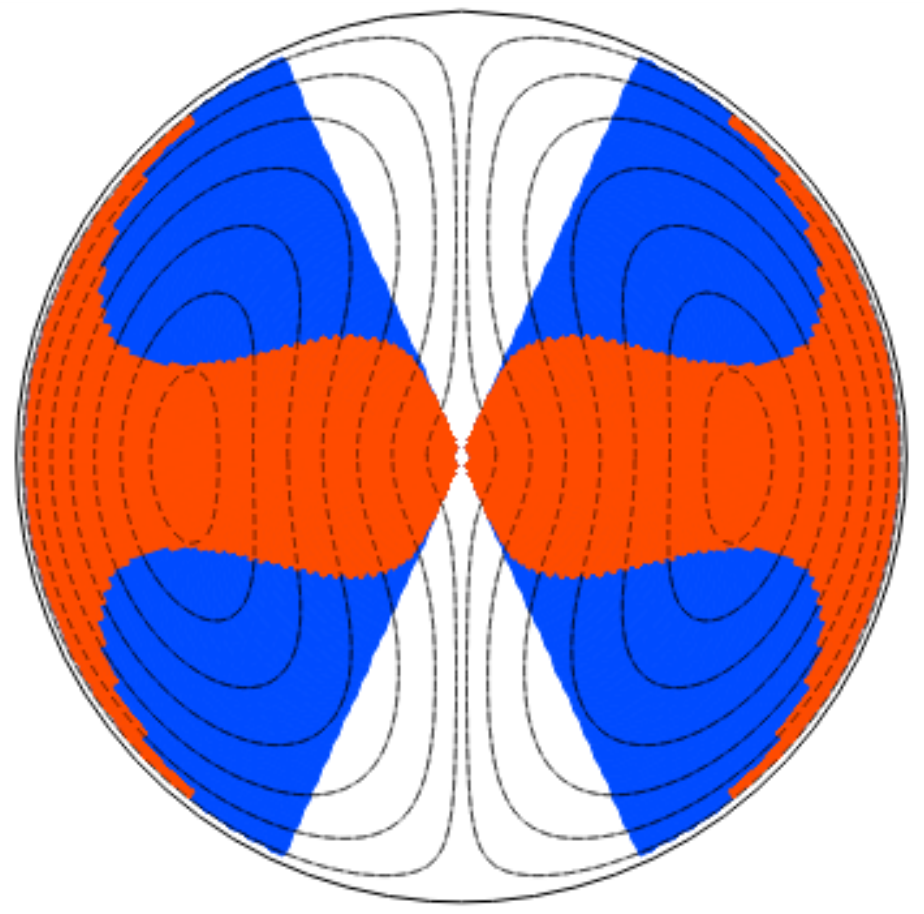}
        \hfill~
    \end{center}
    \caption{\textbf{Left:} Two examples of trapped gravito-magnetic waves at low frequency
    in a poloidal field. The two trajectories have different initial
    conditions. Dashed lines represent the magnetic field lines. \textbf{Right:} Propagation zones
    for a magneto-gravito-inertial wave in orange and its gravito-inertial equivalent in blue for
    two different poloidal magnetic field configurations. The black lines are the magnetic field lines.}
    \label{fig:results}
\end{figure}

We computed a large number of trajectories in polytropic stellar models with different magnetic morphologies.
In addition to formerly known behaviors, namely regular trajectories, island chains and chaotic
trajectories, we have found a new one: trapped trajectories, which are trapped along a portion
of a field line while their wavevector diverges. This behavior is observed for toroidal and poloidal fields. 

In the case of a poloidal field, we observe a significant modification of the propagation zones of
sub-inertial gravito-inertial waves that directly depends on the strength and the topology of the
magnetic field. In a more general case, we observe that the propagation zones strongly depend on the
wavevector, which is not the case for pure gravito-inertial waves. We link this observation to the fact that the Alfv\'en frequency $\omega_{\rm A}$ depends on the wavevector, in contrast to both the
Brunt-V\"ais\"al\"a frequency $N_0$ and the Coriolis frequency $f$, which are characteristic of the gravito-inertial waves. 

\section{Conclusion \& perpectives}

In this work we have derived a dispersion relation for magneto-gravito-inertial waves. We then
used it to explore the parameter space with a ray-tracing method, which allowed us to show
the strong influence of the magnetic field on the dynamics of gravito-inertial waves. 
A large amount of data has been produced throughout this work, and new tools could be developed to efficiently analyse them. 
In the long run, this line of work may allow us to link the properties of the internal magnetic field
to observable quantities such as oscillation frequencies. 
Although it is a challenging issue, it would be interesting to derive a more general dispersion relation taking the derivatives of background quantities into account.
This would make the near-surface dynamics of waves more accurate.
Moreover, in reality, magneto-gravito-inertial waves may escape the star and follow external field lines.
This could be studied with the tools used in this work.

\begin{acknowledgements}
    The authors acknowledge support from ERC SPIRE (grant number 647383) and from PLATO CNES grant at CEA-Saclay.
    A.V. was supported by ENS Lyon and CEA-Saclay.
\end{acknowledgements}

\bibliographystyle{aa}  
\bibliography{valade}

\end{document}